\documentclass[fleqn,10pt]{wlscirep}
\usepackage[utf8]{inputenc}
\usepackage[T1]{fontenc}
\usepackage{lineno}
\usepackage{multirow}
\usepackage{gensymb}
\linenumbers

\title{Full scale, microscopically resolved tomographies of sandstone and carbonate rocks augmented by experimental porosity and permeability values}

\author[1,$\dag$, *]{Matheus Esteves Ferreira}
\author[1,$\dag$]{Mariana Del Grande}
\author[1]{Rodrigo Neumann Barros Ferreira}
\author[1]{Ademir Ferreira da Silva}
\author[1]{Márcio Nogueira Pereira da Silva}
\author[1]{Jaione Tirapu-Azpiroz}
\author[2]{Everton Lucas-Oliveira}
\author[2]{Arthur Gustavo de Araújo Ferreira}
\author[3]{Renato Soares}
\author[3]{Christian B. Eckardt}
\author[2]{Tito J Bonagamba}
\author[2]{Mathias Steiner}

\affil[1]{IBM Research, Rio de Janeiro, 20031-170, Brazil}
\affil[2]{University of São Paulo, São Carlos Institute of Physics, São Carlos, 13560-970, Brazil}
\affil[3]{Solintec Consultoria e Serviços de Geologia Ltda,  Rio de Janeiro, 21031-490, Brazil}

\affil[*]{corresponding author(s): Matheus Esteves Ferreira (m.estevesf@ibm.com), Jaione Tirapu-Azpiroz (jaionet@br.ibm.com)}

\affil[$\dag$]{these authors contributed equally to this work}

\begin{abstract}
We report a dataset containing full-scale, 3D images of rock plugs augmented by petrophysical lab characterization data for application in digital rock and capillary network analysis. Specifically, we have acquired microscopically resolved tomography datasets of 18 cylindrical sandstone and carbonate rock samples having lengths of 25.4 mm and diameters of 9.5 mm, respectively. Based on the micro-tomography data, we have computed porosity-values for each imaged rock sample. For validating the computed porosity values with a complementary lab method, we have measured porosity for each rock sample by using standard petrophysical characterization techniques. Overall, the tomography-based porosity values agree with the measurement results obtained from the lab, with values ranging from 8\% to 30\%. In addition, we provide for each rock sample the experimental permeabilities, with values ranging from 0.4 mD to above 5D. This dataset will be essential for establishing, benchmarking, and referencing the relation between porosity and permeability of reservoir rock at pore scale.

\end{abstract}
\begin{document}
\nolinenumbers
\flushbottom
\maketitle

\thispagestyle{empty}

\section*{Background \& Summary}

The use of X-ray micro-computed-tomography ($\mu$CT) has transformed the study of porous media such as reservoir rocks. Extracted from high-resolution 3D images, the spatial distribution, geometry, and morphology of the pore space is now being used as a basis for computational fluid dynamics simulations and for estimating physical properties such as porosity and permeability. In rock samples most of the pores have diameters of the order of micrometers or below. However, the rock samples, typically cylindrical in shape and referred to as “plugs”, have a dimension in the centimeter range. As a result, a trade-off exists between the overall sampled volume of the rock plug and the microscopic resolution that can be achieved. Consequently, the literature predominantly reports either low-resolution, i.e. 10-100 $\mu$m/voxel, studies of large plugs with diameters of 10-50mm, or, alternatively, high-resolution, i.e. 1-10 $\mu$m/voxel, studies of smaller plugs with diameters of 1-10 mm \cite{1,2,3,4,5,6}.

Laboratory measurements of a rock’s porosity and permeability are routinely performed on plugs having a diameter and a height of 38.1mm, respectively. This leads to a substantial mismatch, often more than 1000-fold, between the sample volumes that are imaged and probed in lab measurements, respectively. The difference in scales complicates the comparison between the porosity and permeability values obtained from direct petrophysical measurements with those indirectly measured from $\mu$CT images. Such an analysis can be performed for spatially homogeneous rock samples such as sandstones \cite{7}, however, it might fail for rather inhomogeneous rock samples, such as carbonates.

In the case of $\mu$CT studies, the porosity and permeability measurements are computed from the generated 3D volumes either through calculation of the void space or through fluid simulations, respectively. To distinguish these measurements from the petrophysical characterization, we refer to these calculated porosity values as “computed” values and the direct petrophysical characterization as “laboratory” measurements. 

In this work, we report full-scale, microscopically resolved X-ray tomographies of rock samples having the shape of a cylindrical plug with a diameter of 9.5 mm and a height of 25.4 mm. Each rock tomography is augmented by porosity and permeability values which were independently measured on the same rock samples in the lab. All rock samples were imaged and analyzed by following the same data acquisition protocol and by using the same equipment. 

\begin{figure}[h!]
\centering
\includegraphics[width=\linewidth]{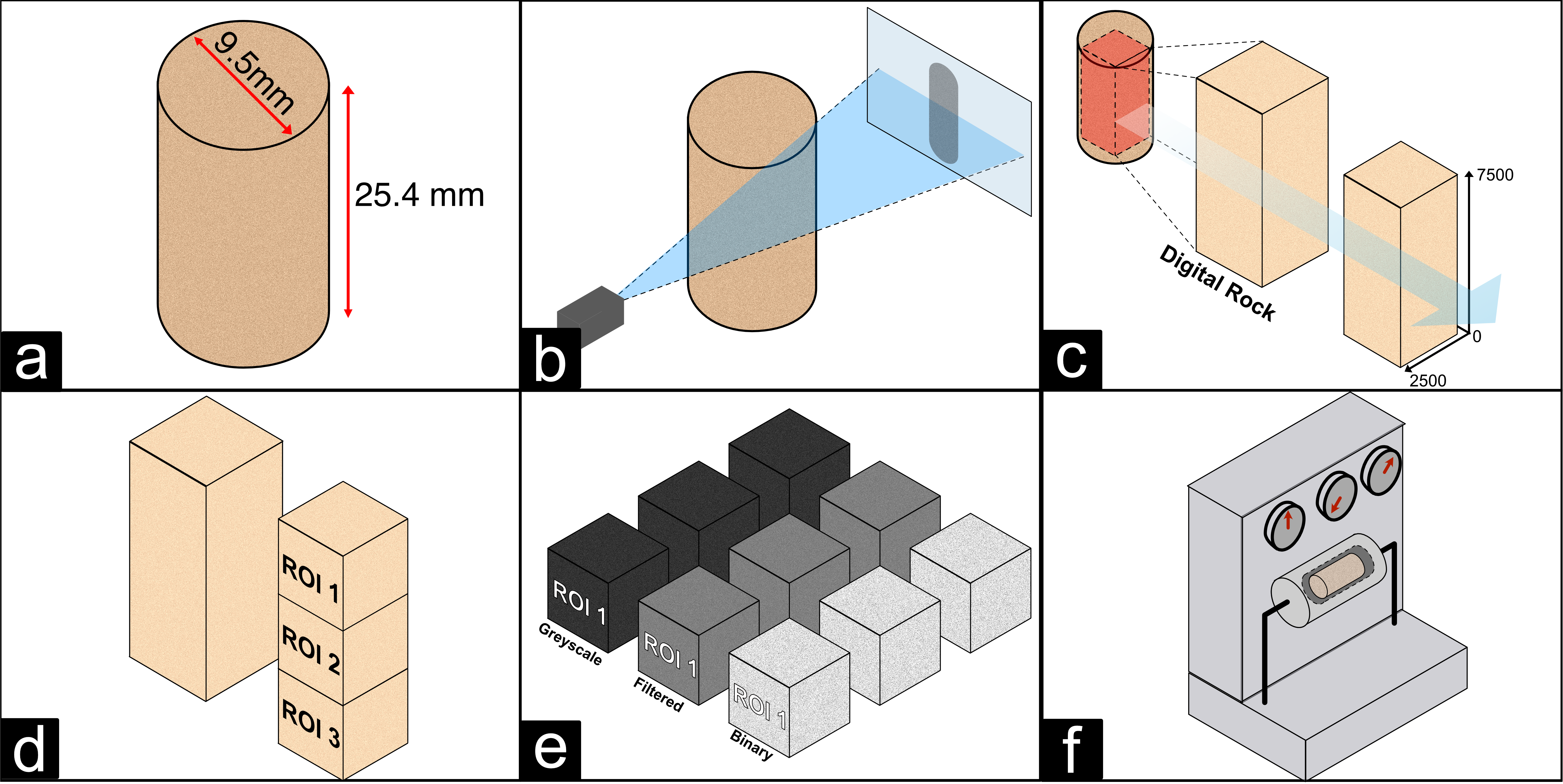}
\caption{\textbf{Conceptual overview of the rock sample study.} (a) Schematic of a cylindrical rock plug sample having a length of 25.4 mm and a diameter of 9.5 mm. (b) Schematic of the X-Ray $\mu$CT imaging process. (c) Visualization of the image cube cropping process. (d) Data cube subdivision by regions of interest (ROI). (e) Data cube processing from greyscale to binary images. (f) Schematic representation of porosity and permeability measurements in the lab.}
\label{fig:1}
\end{figure}

Figure \ref{fig:1} illustrates a schematic overview of the steps followed in this study to produce the digital rock tomography dataset. As depicted in Figure \ref{fig:1}b-\ref{fig:1}d, for each rock sample in the dataset, the scanned image is provided in three different formats: a first image file in raw format of the largest inscribed parallelepiped within the plug, a second raw file where the original image is cut to conform to a standardized parallelepiped of size $2500\times2500\times7500$ voxels, and, lastly, a set of three $2500^3$ voxel cubes extracted from the standardized image. Finally, Figure \ref{fig:1}e illustrates the measurement of porosity and permeability of each sample in the lab. In the following, we discuss the methods involved in image data acquisition and post-processing as well as the laboratory measurement techniques for obtaining porosity and permeability values.

\section*{Methods}

\subsection*{Rock Plug Sample Description}

The carbonate and sandstone rock plug samples (Kokurec Industries Inc.) have a size of 9.5 mm diameter and 25.4 mm in length, as shown in Figure \ref{fig:1}a. The sample size was chosen for enabling full scale imaging with high resolution and petrophysical characterization on the same sample. Table \ref{Table:1} lists all rock samples analyzed in this work.

\subsection*{Rock Sample Imaging and Tomography}

\begin{figure}[h!]
\centering
\includegraphics[width=\linewidth]{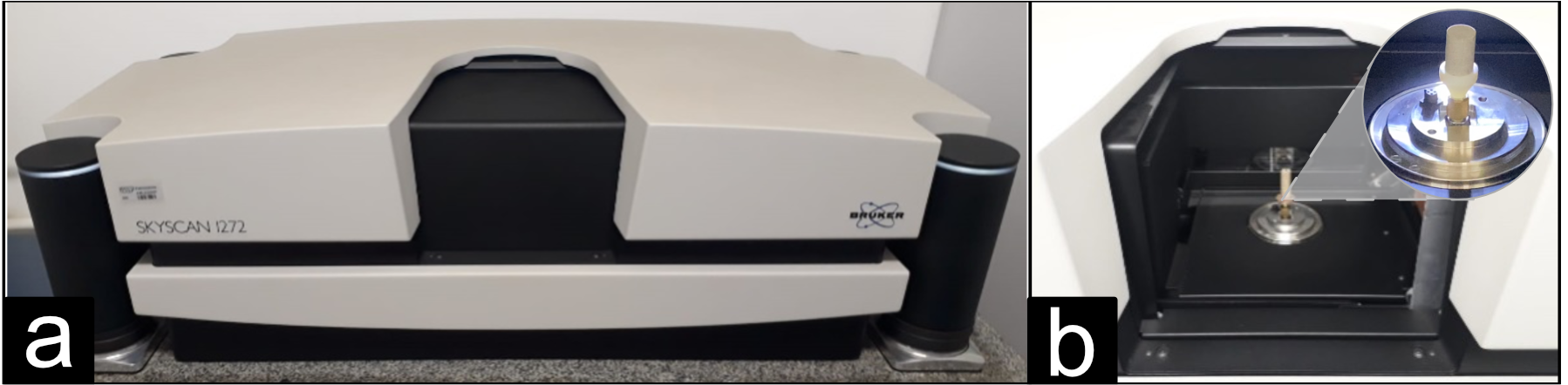}
\caption{\textbf{Experimental setup for rock micro-tomography.} (a) X-Ray $\mu$CT System. (b) Rotational sample stage with mounted rock plug sample. }
\label{fig:2}
\end{figure}

We have acquired digital 3D image volumes from all samples in Table \ref{Table:1} using the X-ray $\mu$CT system (Skyscan 1272, Bruker) shown in Figure \ref{fig:2}a. During image acquisition, the $\mu$CT system produces a series of two-dimensional projections of the porous rock that are computationally transformed into 3D digital representation. Figure \ref{fig:2}b shows the cylindrical rock plug vertically placed in the $\mu$CT system.

We configured the image acquisition software (SkyScan1272 Control Program, version 1.2.0.0; Bruker) as follows: \textit{I=100 $\mu$A; V=100 kV; Frame Averaging = 3; Cu 0.11 mm filter; Pixel Size = 2.25 $\mu$m; Rotation of 360\degree with 0.1\degree  steps; random movement range = 2 to 4} (exact parameters for each sample are provided in the supplementary information \ref{Sup:1}). 

\begin{figure}[h!]
\centering
\includegraphics[width=\linewidth]{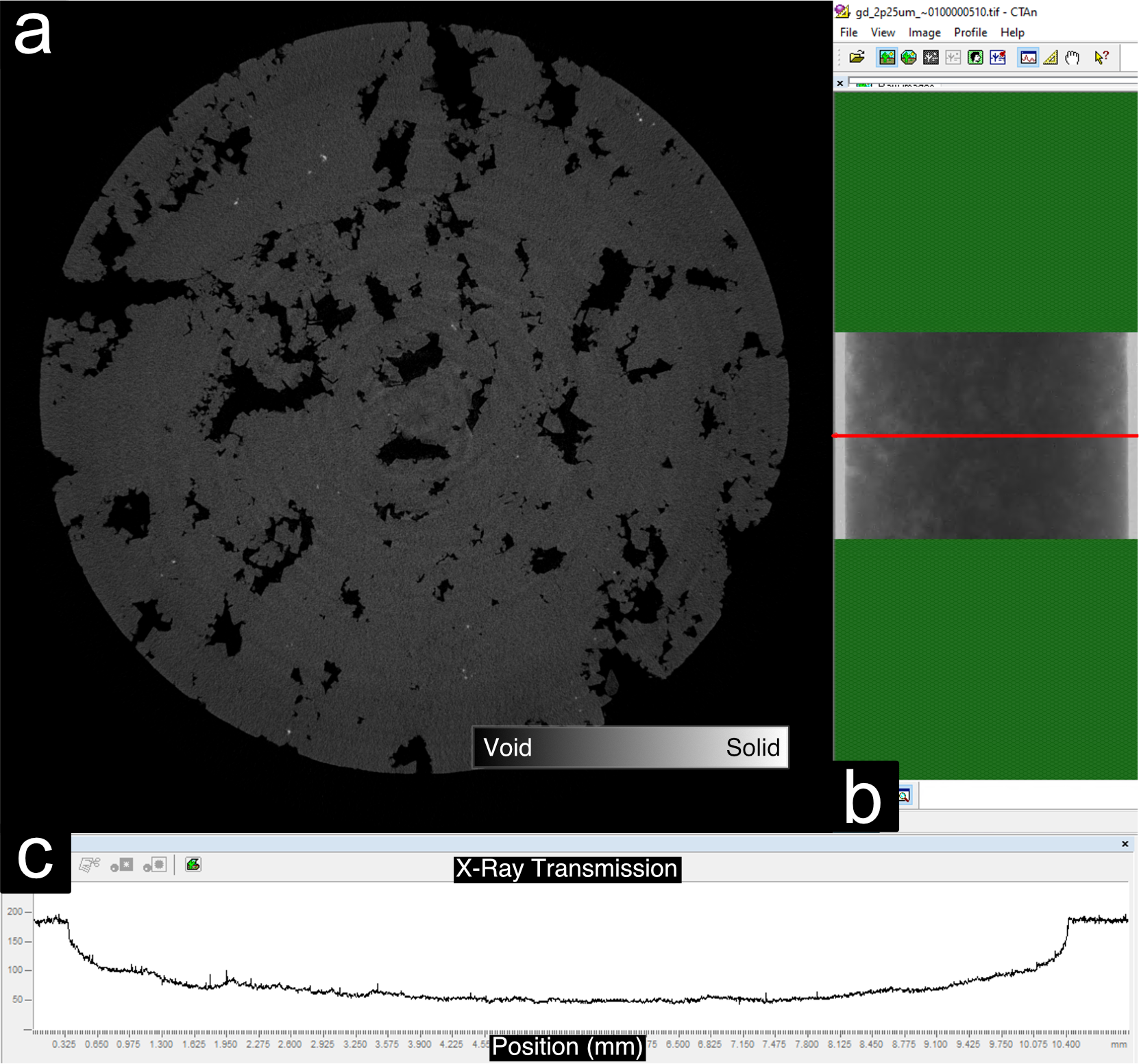}
\caption{\textbf{Analysis of X-ray attenuation through the sample.} (a)  Rock image taken close to the sample center where the darker regions represent the void spaces (b) User interface showing the cross-sectional intensity variations across the center of the sample (in red). (c) Intensity profile along the red line in (b) with a maximum and minimum signal around 200 and 50 grayscale levels, respectively.}
\label{fig:3}
\end{figure}

To ensure a suitable sample size and gauge the X-Ray attenuation through the sample, we computed profile curves along the center of the plug. Figure \ref{fig:3}a  displays a center slice of a digital rock sample, Figure \ref{fig:3}b shows the data acquisition user interface indicating the height of the cross-sectional cutline across the center of the sample (in red), and Figure \ref{fig:3}c shows the transmission intensity profile for the sample GD (Guelph Dolomite). In this example, we observed that the transmission along the sample reaches a minimum grayscale level of around 50 at the sample center, with a maximum value of around 200. Samples with X-Ray transmission close to a 0 were discarded from the study.

\subsection*{Rock Image Data Processing Workflow}

After completion of image data acquisition, the reconstruction of the 3D image was performed by calculating the orthogonal slices from the radial projections using the Feldkamp algorithm \cite{8} implemented within the measurement system software (NRecon, version 1.7.4.6, with the Reconstruction engine InstaRecon, version 2.0.4.6, Bruker). In addition, the reconstruction involves the application of various data processing methods to reduce image artifacts generated by noise in the X-Ray signal during image acquisition. Such signal variations can occur due to fluctuations in the X-ray emission intensity, the detector sensitivity, or through attenuation of lower energy components within denser sample volumes. 

The parameters for the reconstruction include Smoothing (using Gaussian kernel), Ring Artifacts Reduction and Beam-Hardening. We selected the most suitable configuration parameters by scanning the possible values with large steps of trial reconstructions, followed by fine tuning with smaller steps until the result was acceptable. We left the reconstruction histogram unchanged to cut and rescale it uniformly in subsequent steps of data processing. We defined the ROI such that it was contained inside the sample through all the slices. We left the undersample option unchecked as no digital binning was used in this study.

\begin{figure}[h!]
\centering
\includegraphics[width=\linewidth]{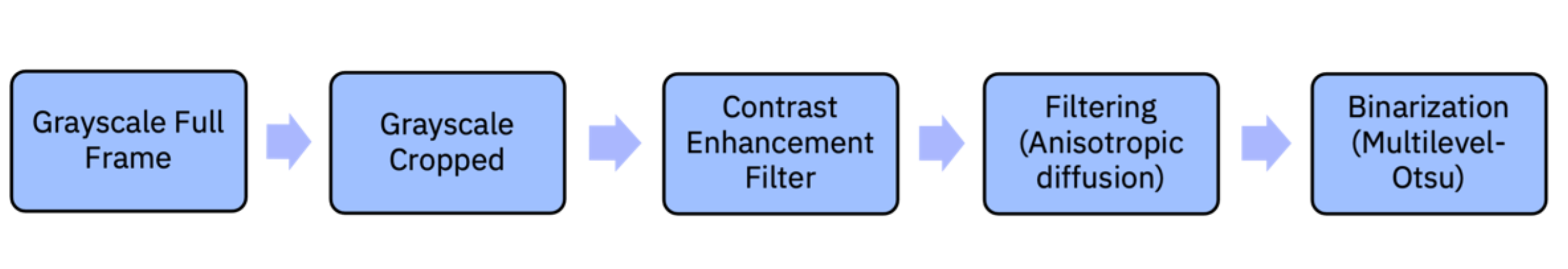}
\caption{\textbf{Rock data processing workflow applied to each image cube.}}
\label{fig:4}
\end{figure}

\begin{figure}[h!]
\centering
\includegraphics[width=\linewidth]{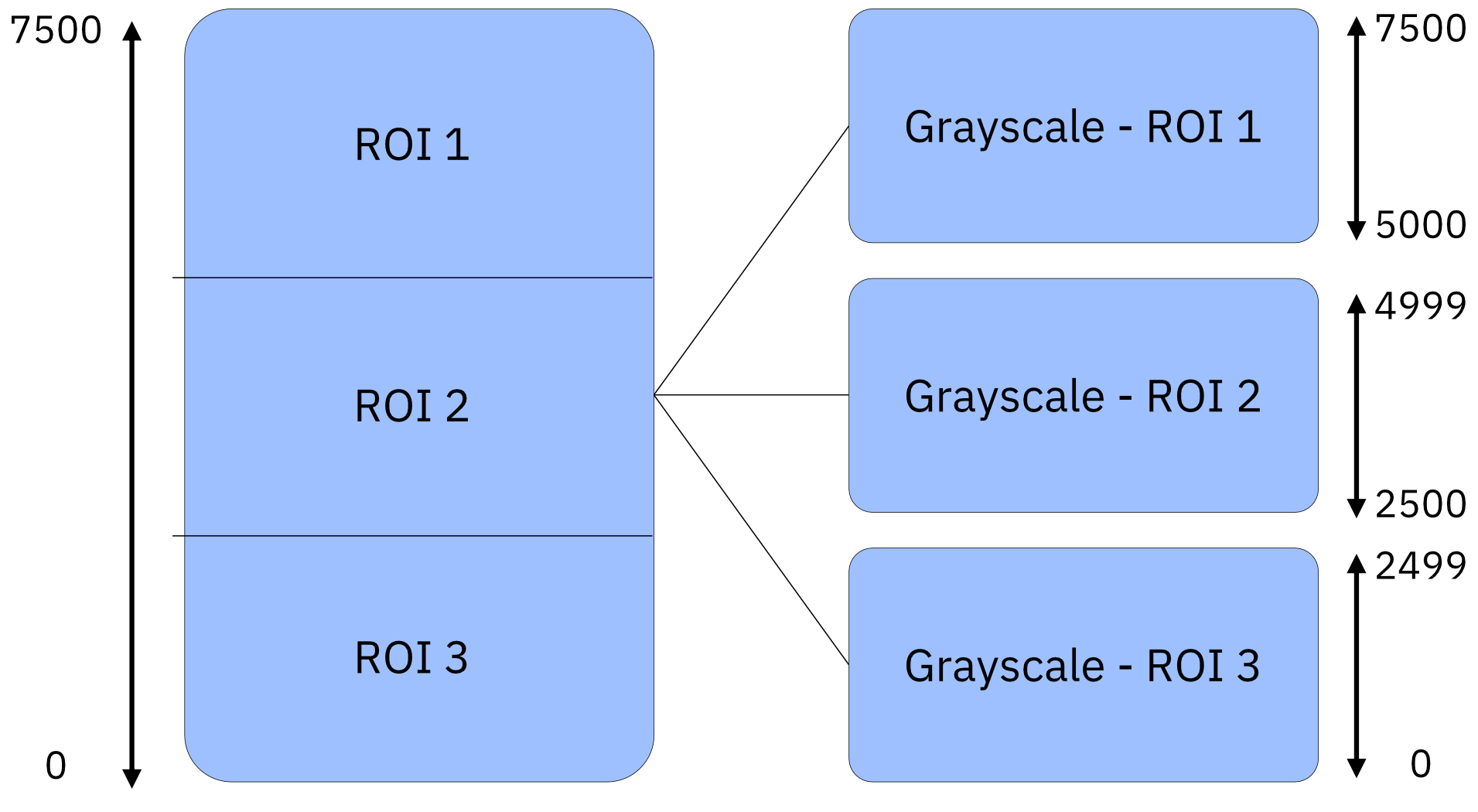}
\caption{\textbf{Splitting of the standardized image data volume into three regions of interest.}}
\label{fig:5}
\end{figure}

Once the 3D digital grayscale rock images were reconstructed, we applied the image data processing workflow outlined in Figure \ref{fig:4} for removing measurement artifacts and separate the pore space from the rock matrix. In a first step, we cropped the full digitalized volume obtained from the $\mu$CT measurements to a standard size of $2500\times2500\times7500$ voxels. This way, the image data parallelepiped could be further split equally into three $2500^3$ voxel sized cubes for improved data handling, see Figure \ref{fig:5}.

In a next step, we applied a contrast enhancement filter to account for the varying mineralogic compositions of the samples studied for equalizing the contrast across all image data sets. The filter was applied to each $2500^3$ voxel volume independently, cutting off the histogram at the grayscale level in which the accumulated histogram achieved 99.8\%, and mapping the remaining grayscale levels back to the [0, 255] interval, thus ensuring an efficient utilization of the entire gray level range.

In a next step, the image data was processed by an anisotropic diffusion filter implemented within the measurement system software (Bruker, version 1.20.8.0) for reducing image noise. The filter was set to 3D space, the type used was Privilege high contrast edges (Perona-Malik), the number of iterations set to 5 and the gradient threshold set to 10. The user defined integration constant option was left unchecked.

Finally, we evaluated both Multi-Otsu and Otsu methods \cite{9,10} for determining a grayscale threshold level for segmentation into solid and void spaces, leading to a binary cubic volume. We observed that a binary segmentation was not capable of properly discerning between matrix and pore structure for all samples studied, mainly due to sample sub-porosity, i.e. image regions of intermediary grayscale levels caused by heterogenous mineral composition, or limited pixel resolution. Therefore, a 3-level Otsu method was chosen.

\begin{figure}[h!]
\centering
\includegraphics[width=\linewidth]{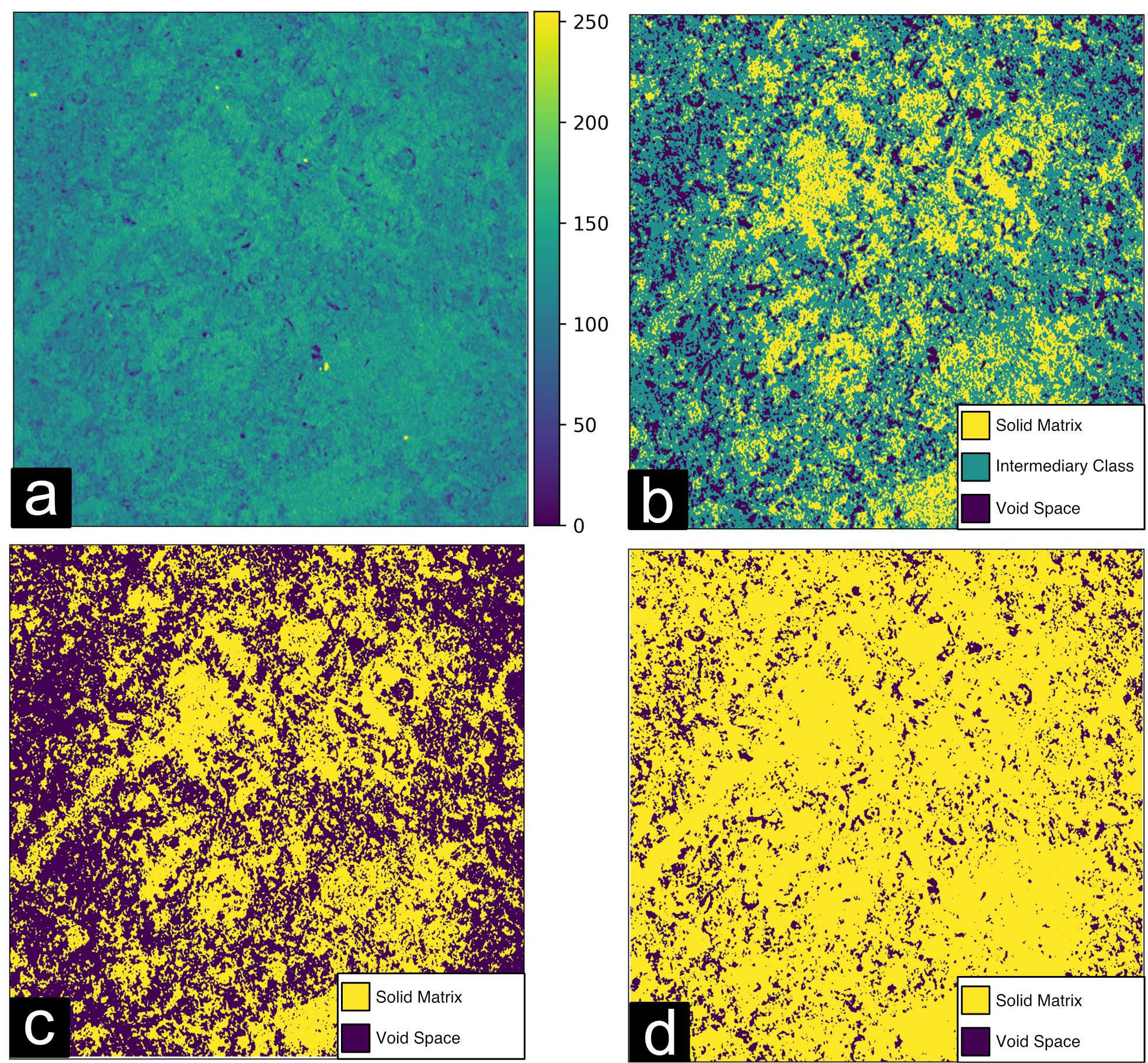}
\caption{\textbf{Effect of the segmentation algorithm on computed porosity.} (a) Filtered grayscale image from sample 5A. (b) Processed image segmented by means of the Multi-Otsu algorithm (n=3) shows three distinct phases. (c) Segmented image using the Otsu algorithm, and (d) segmented image using the Multi-Otsu method after merging the two levels (solid matrix and intermediary class) with the highest values into one representing the solid rock matrix.}
\label{fig:6}
\end{figure}

To ensure proper segmentation, the intermediary class identified by the Multi-Otsu algorithm (corresponding to the sub-porous region) was considered part of the mineral matrix. Figure \ref{fig:6} shows the effect on the digitalized rock image when applying the Multi-Otsu algorithm. Figure \ref{fig:6}a displays the grayscale filtered image extracted from sample 5A after undergoing the various processing steps shown in Figure \ref{fig:4}.  Figure \ref{fig:6}b shows the same rock sample image after the Multi-Otsu algorithm has identified three different regions in this heterogeneous sample, a black region representing the pore space, a yellow area representing the rock matrix and, in light green, the intermediary phase. 

\begin{figure}[h!]
\centering
\includegraphics[width=\linewidth]{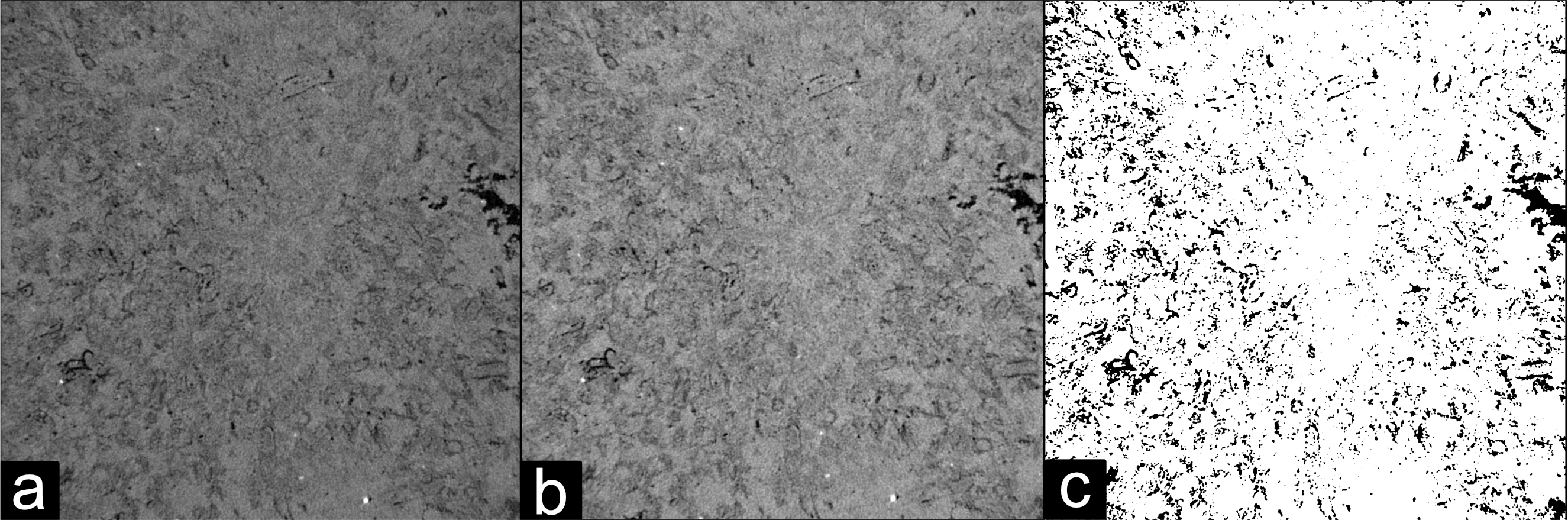}
\caption{\textbf{Representative example of the image processing end-to-end.} Example image in a) raw, b) filtered and c) segmented mode, representing each step in our image processing workflow. Each image has 2500 voxels side length.}
\label{fig:7}
\end{figure}

By calculating the ratio of void to solid space in the binarized volumes, we can estimate the porosity of the sample and compare it with the laboratory measurement value of 13.89\%. We obtain a porosity of 33.68\% with the 2-level Otsu method while the 3-level Multi-Otsu method provides 8.5\% porosity (after merging two levels in the ROI 1 cube). Although this approach may lead to sub-estimation of porosities from the $\mu$CT images, it helped to mitigate the limitations due to the lack of region contrast and produced more accurate porosity estimates across all samples. Despite these limitations, we expect that, due to the resolution limit of the X-Ray $\mu$CT, porosity estimates based on the tomographic volumes yield values lower than those obtained in the petrophysical characterization, which seems compatible with our results. Table \ref{Table:2} lists for each sample the thresholds applied in this study. The cutoff point for binarization was defined as setting pixels equal or greater than the value of the threshold to 1. As a representative example of the effects of data processing, we show in Figure \ref{fig:7} a single tomographic slice in raw, filtered, and binary formats, respectively.

\subsection*{Lab Experimental Characterization of Petrophysical Properties of Rock Samples}

After image data acquisition, we measured porosity and absolute permeability of each rock sample at an overburden pressure of 500 psi in Nitrogen gas at 21ºC using standard equipment (UltraPore-300 and UltraPerm-600, Core Labs). We determined pore and solid volumes based on the known flow cell volume and overburden pressure by assuming isothermic conditions. We estimated the pore density from the ratio between the solid mass and volume. All petrophysical characterization methods were performed following API RP 40 best practices for core analysis \cite{11}. The experimental porosity and permeability values are provided in Table \ref{Table:3}.

\section*{Data Records}

\begin{figure}[h!]
\centering
\includegraphics[width=\linewidth]{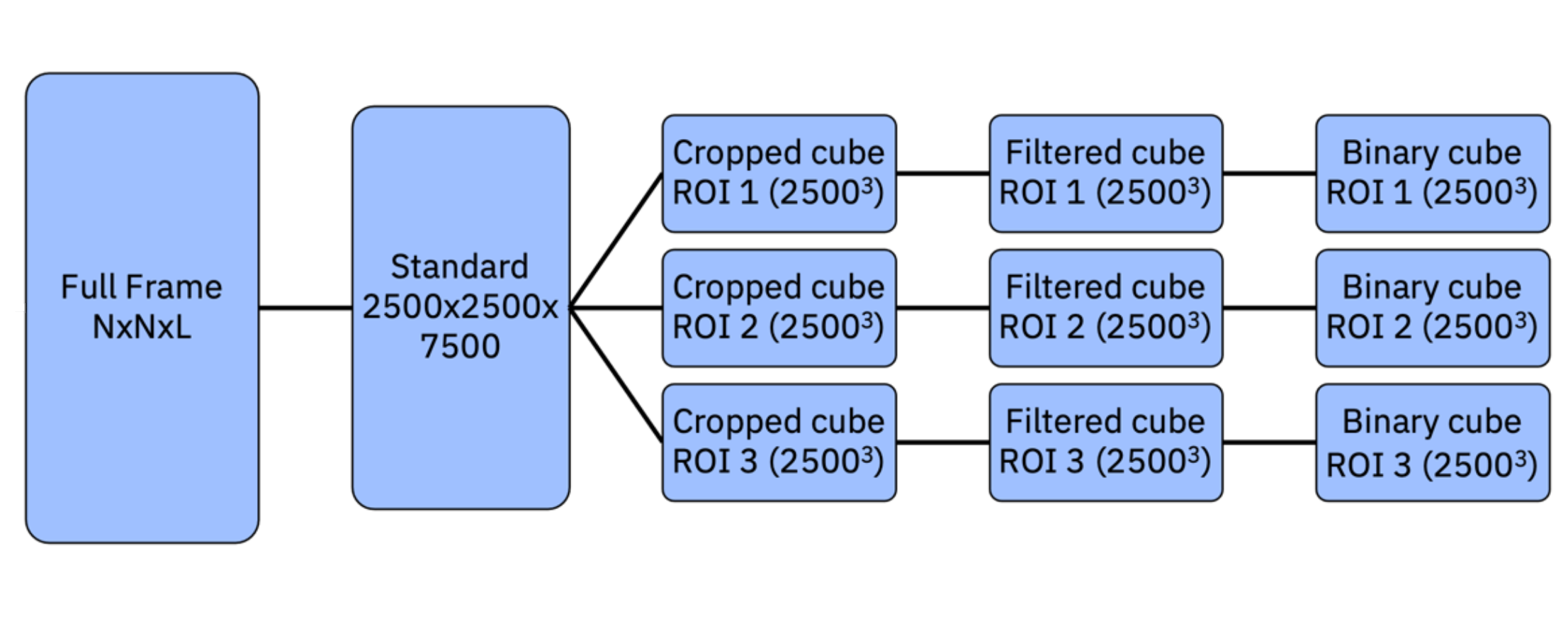}
\caption{\textbf{Overview of the dataset file structure.}}
\label{fig:8}
\end{figure}

The dataset \cite{Dataset} is provided in five different volume types and formats for each sample, as summarized in Figure \ref{fig:8}. The suffix inside the parenthesis designates the naming scheme used for the dataset files:

\begin{itemize}
\item \textbf{Full Frame (\_grayscale\_full):} Data obtained from the reconstruction of the $\mu$CT projections. During reconstruction, the volume edges are removed, however, the largest inscribed parallelepiped within the plug is retained, thus leading to different sized parallelepipeds.

\item \textbf{Standard (\_grayscale\_standard):} Volume cropped into a standard size of $2500\times2500\times7500$ voxels.

\item \textbf{Cropped cubes (\_grayscale\_ROI-X):} $2500^3$ voxel cubes extracted from the standard volume. The X designates the number of the cube, with values ranging from 1 to 3, cut top-down from the parallelepiped. 

\item \textbf{Filtered cubes (\_grayscale\_filtered\_ROI-X):} Data obtained from the grayscale cubes through the application of contrast enhancement and noise reduction filters. The X designates the number of the cube, with values ranging from 1 to 3, cut top-down from the parallelepiped. 

\item \textbf{Binarized cubes (\_binary\_ROI-X):} Binary image data obtained from the filtered grayscale cubes. Each grayscale cube was segmented at threshold level (see Table \ref{Table:2}) calculated using the Multi-Otsu algorithm with a number of classes set to three. The X designates the number of the cube, with values ranging from 1 to 3, cut top-down from the parallelepiped. 

\end{itemize}

In addition to the above, we provided as supporting information:
\begin{itemize}
    \item \textbf{HDR file:} File containing the cube size information for each sample.
\end{itemize}

The dataset \cite{Dataset} acquired  in  this  study  and  reported  in  the  manuscript  is  available  under the DOI:

\href{https://doi.org/10.25452/figshare.plus.21375565}{10.25452/figshare.plus.21375565}.

\section*{Technical Validation}

\subsection*{Comparison between computed and laboratory porosities}

We now compare the porosity computed based on the rock image data (with pixel values ranging from 0 to 1 for void and solid matrix spaces, respectively) with the porosities measured following standard petrophysical lab methodology. For each ROI cube, we computed the porosity based on eq. \ref{eq:1}:
\begin{equation}
    Porosity = 1 - Mean(ROI)
\label{eq:1}
\end{equation}
The mean porosity value of each sample was calculated by averaging the porosity values obtained for ROI 1, 2, and 3. 

\begin{figure}[ht]
\centering
\includegraphics[width=\linewidth]{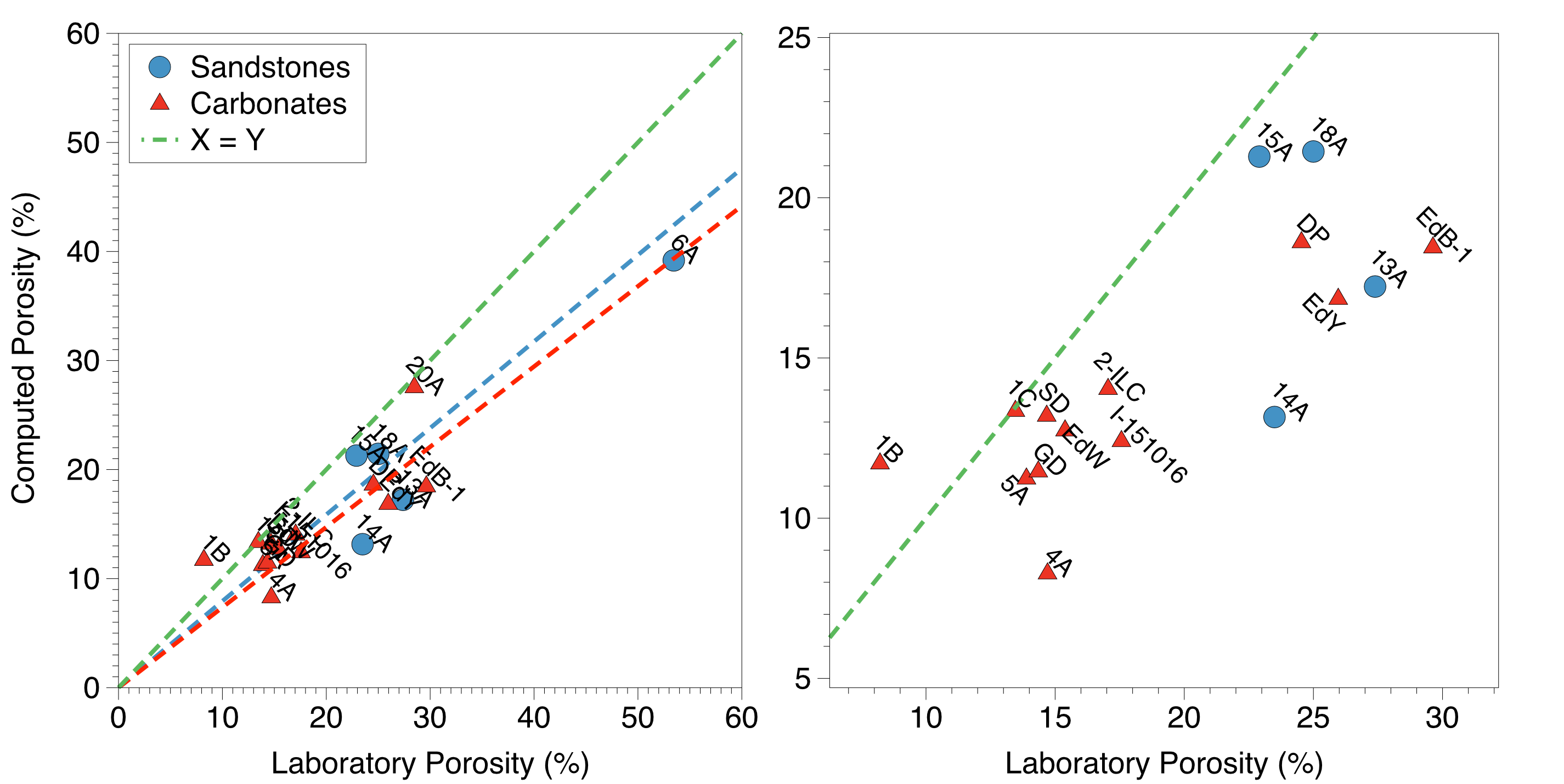}
\caption{\textbf{Analysis of computed and laboratory porosities for sandstone (blue) and carbonate (red) samples.} The green line represents identity. The blue ($R^2$ = 0.243) and red ($R^2$ = 0.889) lines represent linear fits to the carbonate and sandstone data, respectively. The panel on the right represents a zoom of the plot on the left.}
\label{fig:9}
\end{figure}

Figure \ref{fig:9} compares the computational (averaged between all three ROIs) and laboratory porosity results for all samples in the dataset. As expected, except for sample 1B, all samples are located close to or below the green line due to under-estimation of porosity, most probably caused by limitations in image resolution. Overall, we find that the image-based method provides robust porosity estimates for both sandstone and carbonate samples. Future research work is needed to connect the porosity and permeability values for each sample based on image analysis. To that end, we believe that the data published in this study provides key contributions.

\section*{Code availability}

The algorithms used for processing and segmenting the raw grayscale images are available as Python code at: 

\href{https://github.com/IBM/microCT-Dataset}{https://github.com/IBM/microCT-Dataset}

The code repository contains Jupyter Notebooks for simplifying data processing and visualization along with usage guidance.

\bibliography{References}

\section*{Acknowledgements}

The authors acknowledge project support by Bruno Flach and Alexandre Pfeifer (both IBM). TJB acknowledges the support of the following Brazilian Institutions: University of São Paulo (USP), Petróleo Brasileiro S.A. (Petrobras/CENPES, 2020/00010-0), and National Council for Scientific and Technological Development (CNPq, 308076/2018-4).

\section*{Author contributions statement}

R.N.B.F, E.L.O, A.A.F, R.S., C.B.E, T.J.B, and M.S. conceived the study. M.E.F, M.D.G, R.N.B.F, A.F.S, E.L.O, A.A.F, R.S., C.B.E, T.J.B, and M.S. designed the experiments. M.D.G and R.N.B.F acquired the $\mu$CT data. R.S and C.B.E provided the petrophysical characterizations. M.E.F, R.N.B.F, M.N developed code for processing and analysing the $\mu$CT data. M.E.F, M.D.G, and M.N processed and analysed the $\mu$CT data. M.E.F, M.D.G, R.N.B.F, J.T.A., E.L.O, A.A.F, T.J.B, and M.S reviewed, discussed and curated the dataset. M.E.F, M.D.G, and J.T.A externalized the data and the code generated in this work. M.E.F, M.D.G, R.N.B.F, A.F.S, M.N., J.T.A, E.L.O., A.A.F, T.J.B, and M.S wrote and edited the manuscript. All authors reviewed and approved the manuscript. 

\section*{Competing interests}

The authors declare that they have no known competing financial interests or personal relationships that could have appeared to influence the work reported in this paper.

\section*{Figures \& Tables}


\begin{table}[h!]
\centering
\begin{tabular}{|l|l|l|}
\hline
\textbf{Sample Name} & \textbf{Rock} & \textbf{Type} \\
\hline
1B & Silurian Dolomite & Carbonate \\
\hline
1C & Silurian Dolomite & Carbonate \\
\hline
4A & Indiana Limestone & Carbonate \\
\hline
5A & Lueders & Carbonate \\
\hline
6A & Mr. Gambier & Carbonate \\
\hline
13A & Castlegate & Sandstone \\
\hline
14A & Carbon Tan & Sandstone \\
\hline
15A & Bentheimer & Sandstone \\
\hline
18A & Liver Rock & Sandstone \\
\hline
20A & Idaho Gray & Sandstone \\
\hline
SD & Silurian Dolomite & Carbonate \\
\hline
I-151016 & Indiana Limestone & Carbonate \\
\hline
GD & Guelph Dolomite & Carbonate \\
\hline
EdY & Edwards Yellow & Carbonate \\
\hline
EdW & Edwards White & Carbonate \\
\hline
EdB-1 & Edwards Brown  & Carbonate \\
\hline
DP & Desert Pink  & Carbonate \\
\hline
2-ILC & Indiana Limestone  & Carbonate \\
\hline
\end{tabular}
\caption{\textbf{List of rock samples analyzed in this study}}
\label{Table:1}
\end{table} 

\begin{table}[h!]
\centering
\begin{tabular}{|l|l|l|l|}
\hline
\textbf{Sample Name} & \textbf{ROI 1 Threshold} & \textbf{ROI 2 Threshold} & \textbf{ROI 3 Threshold} \\
\hline
1B & 93 & 93 & 98 \\
\hline
1C & 91 & 93 & 94 \\
\hline
4A & 94 & 95 & 97 \\
\hline
5A & 105 & 120 & 126 \\
\hline
6A & 79 & 79 & 81 \\
\hline
13A & 72 & 73 & 74 \\
\hline
14A & 100 & 102 & 103 \\
\hline
15A & 67 & 74 & 77 \\
\hline
18A & 52 & 54 & 59 \\
\hline
20A & 76 & 77 & 80 \\
\hline
SD & 83  & 83  & 85 \\
\hline
I-151016 & 88 & 88 & 90 \\
\hline
GD & 93 & 93 & 93 \\
\hline
EdY & 88 & 89 & 84 \\
\hline
EdW & 106 & 121 & 124 \\
\hline
EdB-1 & 64 & 68 & 66 \\
\hline
DP & 88 & 82 & 84 \\
\hline
2-ILC & 78 & 80 & 80 \\
\hline
\end{tabular}
\caption{\textbf{Computed thresholds used in the segmentation of each ROI.}}
\label{Table:2}
\end{table} 

\begin{table}[h!]
\centering
\begin{tabular}{|c|c|cccc|cc|}
\hline
\multicolumn{1}{|l|}{\multirow{3}{*}{\textbf{Sample Name}}} & \multicolumn{1}{l|}{\multirow{3}{*}{\textbf{Laboratory Porosity}}} & \multicolumn{4}{l|}{\multirow{2}{*}{\textbf{Computed Porosity}}}                                                                                      & \multicolumn{2}{l|}{\textbf{Laboratory Permeability}}                                 \\ \cline{7-8} 
\multicolumn{1}{|l|}{}                                      & \multicolumn{1}{l|}{}                                                & \multicolumn{4}{l|}{}                                                                                                                                 & \multicolumn{1}{l|}{\textbf{Air}}           & \multicolumn{1}{l|}{\textbf{Klinkenberg}} \\ \cline{3-8} 
\multicolumn{1}{|l|}{}                                      & \multicolumn{1}{l|}{}                                                & \multicolumn{1}{l|}{\textbf{ROI 1}} & \multicolumn{1}{l|}{\textbf{ROI 2}} & \multicolumn{1}{l|}{\textbf{ROI 3}} & \multicolumn{1}{l|}{\textbf{Mean (\%)}} & \multicolumn{1}{l|}{\textbf{(mD)}}          & \multicolumn{1}{l|}{\textbf{(mD)}}        \\ \hline
1B                                                          & 8.22                                                                 & \multicolumn{1}{c|}{0.13}           & \multicolumn{1}{c|}{0.12}           & \multicolumn{1}{c|}{0.11}           & 11.70                               & \multicolumn{1}{c|}{67.39}                  & 61.34                                     \\ \hline
1C                                                          & 13.46                                                                & \multicolumn{1}{c|}{0.15}           & \multicolumn{1}{c|}{0.13}           & \multicolumn{1}{c|}{0.13}           & 13.34                               & \multicolumn{1}{c|}{373.14}                 & 353.53                                    \\ \hline
4A                                                          & 14.71                                                                & \multicolumn{1}{c|}{0.08}           & \multicolumn{1}{c|}{0.08}           & \multicolumn{1}{c|}{0.08}           & 8.27                                & \multicolumn{1}{c|}{23.11}                  & 20.41                                     \\ \hline
5A                                                          & 13.89                                                                & \multicolumn{1}{c|}{0.08}           & \multicolumn{1}{c|}{0.13}           & \multicolumn{1}{c|}{0.12}           & 11.23                               & \multicolumn{1}{c|}{55.79}                  & 50.66                                     \\ \hline
6A                                                          & 53.45                                                                & \multicolumn{1}{c|}{0.38}           & \multicolumn{1}{c|}{0.40}           & \multicolumn{1}{c|}{0.39}           & 39.18                               & \multicolumn{1}{c|}{144.41}                 & 134.37                                    \\ \hline
13A                                                         & 27.39                                                                & \multicolumn{1}{c|}{0.17}           & \multicolumn{1}{c|}{0.17}           & \multicolumn{1}{c|}{0.17}           & 17.23                               & \multicolumn{1}{c|}{638.56}                 & 610.16                                    \\ \hline
14A                                                         & 23.49                                                                & \multicolumn{1}{c|}{0.13}           & \multicolumn{1}{c|}{0.13}           & \multicolumn{1}{c|}{0.13}           & 13.16                               & \multicolumn{1}{c|}{67.56}                  & 61.53                                     \\ \hline
15A                                                         & 22.91                                                                & \multicolumn{1}{c|}{0.22}           & \multicolumn{1}{c|}{0.21}           & \multicolumn{1}{c|}{0.21}           & 21.29                               & \multicolumn{1}{c|}{0.42}                   & 0.31                                      \\ \hline
18A                                                         & 25.00                                                                & \multicolumn{1}{c|}{0.21}           & \multicolumn{1}{c|}{0.21}           & \multicolumn{1}{c|}{0.22}           & 21.44                               & \multicolumn{1}{c|}{392.97}                 & 372.64                                    \\ \hline
20A                                                         & 28.48                                                                & \multicolumn{1}{c|}{0.28}           & \multicolumn{1}{c|}{0.28}           & \multicolumn{1}{c|}{0.27}           & 27.54                               & \multicolumn{1}{c|}{\textgreater{}5,000.00} & 5,000.00                                  \\ \hline
SD                                                          & 14.67                                                                & \multicolumn{1}{c|}{0.14}           & \multicolumn{1}{c|}{0.13}           & \multicolumn{1}{c|}{0.13}           & 13.19                               & \multicolumn{1}{c|}{80.59}                  & 73.80                                     \\ \hline
I-151016                                                    & 17.57                                                                & \multicolumn{1}{c|}{0.13}           & \multicolumn{1}{c|}{0.13}           & \multicolumn{1}{c|}{0.11}           & 12.40                               & \multicolumn{1}{c|}{50.11}                  & 45.23                                     \\ \hline
GD                                                          & 14.35                                                                & \multicolumn{1}{c|}{0.11}           & \multicolumn{1}{c|}{0.12}           & \multicolumn{1}{c|}{0.12}           & 11.45                               & \multicolumn{1}{c|}{1228.80}                & 1184.53                                   \\ \hline
EdY                                                         & 25.96                                                                & \multicolumn{1}{c|}{0.16}           & \multicolumn{1}{c|}{0.17}           & \multicolumn{1}{c|}{0.17}           & 16.84                               & \multicolumn{1}{c|}{11.45}                  & 9.80                                      \\ \hline
EdW                                                         & 15.38                                                                & \multicolumn{1}{c|}{0.12}           & \multicolumn{1}{c|}{0.12}           & \multicolumn{1}{c|}{0.15}           & 12.73                               & \multicolumn{1}{c|}{0.63}                   & 0.48                                      \\ \hline
EdB-1                                                       & 29.63                                                                & \multicolumn{1}{c|}{0.20}           & \multicolumn{1}{c|}{0.16}           & \multicolumn{1}{c|}{0.19}           & 18.45                               & \multicolumn{1}{c|}{14.61}                  & 12.59                                     \\ \hline
DP                                                          & 24.54                                                                & \multicolumn{1}{c|}{0.18}           & \multicolumn{1}{c|}{0.19}           & \multicolumn{1}{c|}{0.19}           & 18.61                               & \multicolumn{1}{c|}{47.98}                  & 43.26                                     \\ \hline
2-ILC                                                       & 17.05                                                                & \multicolumn{1}{c|}{0.15}           & \multicolumn{1}{c|}{0.14}           & \multicolumn{1}{c|}{0.13}           & 14.03                               & \multicolumn{1}{c|}{72.94}                  & 66.57                                     \\ \hline
\end{tabular}
\caption{\textbf{Porosity and permeability values for each rock sample analysed in this study.}}
\label{Table:3}
\end{table} 

\begin{table}[h!]
\centering
\begin{tabular}{|ll|}
\hline
\multicolumn{2}{|l|}{Random Movement} \\ \hline
\multicolumn{1}{|l|}{1C}      & 2     \\ \hline
\multicolumn{1}{|l|}{1B}      & 2     \\ \hline
\multicolumn{1}{|l|}{4A}      & 3     \\ \hline
\multicolumn{1}{|l|}{5A}      & 3     \\ \hline
\multicolumn{1}{|l|}{6A}      & 3     \\ \hline
\multicolumn{1}{|l|}{13A}     & 3     \\ \hline
\multicolumn{1}{|l|}{14A}     & 3     \\ \hline
\multicolumn{1}{|l|}{15A}     & 3     \\ \hline
\multicolumn{1}{|l|}{18A}     & 3     \\ \hline
\multicolumn{1}{|l|}{20A}     & 3     \\ \hline
\multicolumn{1}{|l|}{SD}      & 4     \\ \hline
\multicolumn{1}{|l|}{I15}     & 4     \\ \hline
\multicolumn{1}{|l|}{GD}      & 4     \\ \hline
\multicolumn{1}{|l|}{EdY}     & 3     \\ \hline
\multicolumn{1}{|l|}{EdW}     & 3     \\ \hline
\multicolumn{1}{|l|}{EdB}     & 4     \\ \hline
\multicolumn{1}{|l|}{DP}      & 3     \\ \hline
\multicolumn{1}{|l|}{ILC}     & 4     \\ \hline
\end{tabular}
\caption{\textbf{Random movement parameter used on each sample}}
\label{Sup:1}
\end{table} 

\end{document}